\documentclass[secnumarabic, prb, showkeys]{revtex4}

\begin{document}

\title{The mass and energy of a vapor bubble in a turbulent ideal fluid}
\author{Valery P. Dmitriyev}
\affiliation{Lomonosov University\\
P.O.Box 160, Moscow 117574, Russia}
\email{aether@yandex.ru}
\date{1 February 2006}

\begin{abstract}
 The mass of a bubble in a fluid can be taken as the
mass of the vapor in it. The self-energy of the bubble is defined
as the work performed against the pressure of the fluid in order
to create the bubble.  Taking the vapor to be an ideal gas the
relationship between the self-energy, the mass of the bubble and
the speed of the perturbation wave in a turbulent ideal fluid can
be obtained.
\end{abstract}
 \maketitle

We consider perturbations of the averaged turbulence in an
incompressible  fluid. The speed $c$ of the turbulence
perturbation wave was shown \cite{Troshkin} to be related with the
background level of Reynolds stresses as
\begin{equation}
c^2 = \langle u_1'u_1'\rangle^{(0)} = \langle
u_2'u_2'\rangle^{(0)} = \langle u_3'u_3'\rangle^{(0)}\label{1}
\end{equation}
where $u'_i$ are turbulent fluctuations of the fluid velocity
${\bf u} = \langle\textbf{u}\rangle + {\bf u}'$ and
$\langle..\rangle$ the averaging over a short time interval.

Let a bubble be included into an ideal fluid, and $V^*$ the volume
of the turbulent fluid evaporated into the bubble. The kinetic
energy $K^*$ transferred with the fluid into the gas phase can be
found knowing the volume density of the turbulence energy of the
fluid
\begin{equation}
\frac{1}{2}\varrho (\langle u_1'u_1'\rangle + \langle
u_2'u_2'\rangle + \langle u_3'u_3'\rangle)\label{2}
\end{equation}
where $\varrho$ is the density of the fluid. It should be noted
here that, as a true continuum, the ideal fluid has no heat
energy, or in this system the energy of the turbulence can be
viewed as in a way similar to the heat energy. Calculating $K^*$
for the unperturbed medium we find from (\ref{2}) with the account
of (\ref{1})
\begin{equation}
K^* =  V^*\frac{3}{2}\varrho\langle u_1'u_1'\rangle^{(0)} =
\frac{3}{2}\varrho V^*c^2.\label{3}
\end{equation}
The vapor will be assumed to behave as an ideal gas. The equation
of state of the ideal gas can be written in the form
\begin{equation}
pV = \frac{2}{3}K.\label{4}
\end{equation}
In the mechanical equilibrium the gas pressure $p$ must be equal
to the fluid pressure $\langle p \rangle$. If $V$ is the volume of
the bubble then using (\ref{3}) in (\ref{4}) we get for a bubble
in the unperturbed medium with the background pressure $p_0$:
\begin{equation}
p_0V = \varrho V^*c^2.\label{5}
\end{equation}

The mass of the bubble can be determined from the mass of the gas
contained in it
\begin{equation}
m = \varrho V^*.\label{6}
\end{equation}
The self-energy of a bubble can be defined as the work needed in
order to create the bubble in the unperturbed medium
\begin{equation}
E = p_0V.\label{7}
\end{equation}
Using (\ref{6}) and (\ref{7}) in (\ref{5}) we get
\begin{equation}
E = mc^2.\label{8}
\end{equation}

\end{document}